\documentclass[11pt]{article}
\usepackage[latin1]{inputenc}
\usepackage[english]{babel}
\usepackage{pgf,tikz}
\usetikzlibrary{arrows}
\usepackage{amsmath}
\usepackage{mathrsfs}
\usepackage{amssymb}
\usepackage{amsthm}

\usepackage{epsfig}
\usepackage{graphicx}%

\begin{document}

\title{\Large{\bf \emph{Relaxation Equations: Fractional Models}}}

\author{ {\bf {\large Ester C. F. A. Rosa} \hspace{1cm} {\textbf{E. Capelas de Oliveira}}} \\
 {\small  Department of Applied Mathematics} \\
 {\small IMECC - University of Campinas}\\
 {\small CEP 13083-859, Campinas, SP, Brazil} \\
 {estercps@yahoo.com.br/capelas@ime.unicamp.br }}
\date{}

\maketitle

\thispagestyle{empty}

\begin{abstract}
\noindent The relaxation functions
introduced empirically by Debye, Cole-Cole, Cole-Davidson and Havriliak-Negami
are, each of them, solutions to their respective kinetic equations. In this
work, we propose a generalization of such equations by introducing a fractional
differential operator written in terms of the Riemann-Liouville fractional
derivative of order $\gamma$, $0 < \gamma \leq 1$. In order to solve the
generalized equations, the Laplace transform methodology is introduced and the
corresponding solutions are then presented, in terms of Mittag-Leffler
functions. In the case in which the derivative's order is $\gamma=1$, the
traditional relaxation functions are recovered. Finally, we presente some 2D
graphs of these function.
\end{abstract}


\section{Introduction}

Fractional calculus has greatly developed during the last years and has
established itself as a generalization of classical differential and
integral calculus. Over the last three decades the investigation on this
subject has intensified and applications were found in many fields of science
such as physics, mathematics, chemistry, financial systems, economy and
engineering \cite{Tenreiro:2010}. As examples we may mention the following
applications: the fractional control of engineering systems;  
variational calculus and optimum control for fractional dynamic systems;
numerical and analytical tools and techniques; fundamental explorations of
mechanical, electric and thermal constitutive relations; investigation of
properties of several engineering materials such as viscoelastic polymers,
foams, gels and animal tissues; diffusion
phenomena; bioengineering and biomedical
applications; thermal modeling for engineering systems such as breaks and
tool-machines and, finally, imaging and signal processing
\cite{Hilfer:2000,Kiryakova:1994,Mainardi:2010,Mathai:2010,Oustaloup:1991,
Indranil:2012,Podlubny:1999,Vladimir:2013,Renat:2012}.  

The mathematical investigation of relaxation processes in dielectrics has been
conducted with the use of fractional calculus tools
\cite{Capelas:2011,Haniga:2008,Hilfer1:2002, Mainardi:2014}.  
The properties of dielectric materials are usually represented by the empirical
susceptibility functions as they have been initially modeled by Debye (D) in
1929, then by the Cole brothers (C-C) and, later, in the works of Cole-Davidson
(C-D) and Havriliak-Negami (H-N).\\ 
With the help of the concept of memory developed in the formalism of
Mori-Zwanzig, it is possible to deduce the kinetic equations for the relaxation
processes described by those empirical functions \cite{Popov:2014}. The
equations obtained are expressed in terms of the fractional derivative of
Riemann-Liouville and their solutions are given in terms of Mittag-Leffler
functions \cite{Gorenflo:2014}.

The Mittag-Leffler functions of a real variable $t$, with one, two and three
parameters, are convenient to study anomalous processes such as those occurring
in dielectrics, including the aforementioned classical cases C-C, C-D and H-N.
It is crucial to emphasize that these functions are completely monotonic for
$t>0$ \cite{Capelas:2011,Garrapa:2014}.

In this work we study fractional kinetic equations with a derivative of order 
$\gamma$, $0< \gamma \leq 1$, proposed as generalizations of the classic
kinetic equations associated with relaxation processes in dielectrics, in such
a way that the traditional kinetic equations become particular cases 
of the fractional equations when $\gamma=1$. 

This work is organized as follows: in the first section a brief theoretical
introduction is presented of the four empirical expressions of complex
susceptibility (in the frequency domain) related to Debye's model and to the
anomalous relaxation processes C-C, C-D e H-N.  On the basis of these
expressions of complex dielectric permittivity, we present in the second
section the construction of the kinetic equations for the Debye and the
anomalous relaxation functions. The construction of these kinetic equations is
carried with the introduction of an integral memory function. At the end of the
second section the equations are solved and the relaxation functions (in the
time domain) are written in terms of Mittag-Leffler functions; we also show
their graphic representations. In the third section, the previously mentioned
fractional kinetic equations are solved and the graphs of such solutions are
shown.  Finally, the fourth section, dedicated to the conclusions, presents a
comparative analysis of the results obtained and to the discussion of outline a
proposal for future study.

\section{Classic Empirical Models}

The most important phenomenon associated with a dielectric material is its
polarization, which consists in the change of the distribution of its molecular
and atomic charges when it is subjected to the action of an electric field.
Thus, when an electric field is applied to a dialectric it produces a very
small electric current called dielectric loss, and its constituent particles,
ions or molecules suffer small dislocations or rearrangements, thus altering
their equilibrium positions \cite{Gross:1981}. These molecular parts do not
leave nor reach their state of equilibrium instantaneously: a variable ammount
of time is necessary for this change of positions to take place. This time
lapse necessary for the material to respond to the electric field applied is
called relaxation time. Thus, when submitted to an electric excitation, the
dielectric will respond to this action in an attempt to reestablish its
equilibrium during and after the electric stimulus. The polarization of each
dielectric material depends on the nature of its molecular and atomic chemical
bonds, and there is presently no universal model which can explain the
polarization phenomenon in all materials \cite{Bottcher:1978}.  Debye's
response function was the first theoretical model for the dielectric behavior
of some substances \cite{Debye:1929}.  However, due to its limitations, this
model is incapable of describing in detail the dielectric response of a large
number of solids and liquids. 

Therefore, in the years following the appearance
of Debye's theory, several other response functions were proposed in the
literature to serve as models for describing the dielectric relaxation of many
materials. Among them, we here focus our attention on the model proposed by the
Cole brothers \cite{ColeCole:1941,Cole:1942},  and the one by Cole-Davidson
\cite{Davidson:1950,DavidsonCole:1951}, both of which emerged in attempts to
adjust the response function to the experimental behavior of some dielectric
materials.  There is also the model of H-N \cite{Negami:1966,Havril:1967} which 
can be considered a generalization of the
latter models.  These last three models, called anomalous models, are the most
relevant in the literature; however, there are other models which approach the
phenomena from a different perspective, as the models by Weron and
collaborators \cite{Weron1:2002,Weron2:2008,Weron4:2010,Weron3:2007}, Hilfer
\cite{Hilfer1:2002,Hilfer2:2002}, Hanyga and Seredynska \cite{Hanyga:2008}. 

During the last years, some of these authors and others not mentioned above
have addressed the anomalous relaxation processes of types C-C, C-D and H-N
with the help of fractional calculus, for example, by associating the response
functions to the Mittag-Leffler functions (which turn out to be related with
the complex susceptibility $s=i\omega$ through the Laplace transform) 
\cite{Capelas:2011}; by modeling relaxation processes through equations with
fractional derivatives, whose solutions are also given in terms of
Mittag-Leffler functions; and establishing other theoretical connections
discovered through the use of fractional calculus tools.\\  

\noindent In 1929, Debye conceived a simple model for the relaxation process in which he
supposed a unique relaxation time for all molecules, obtaining the following
expression \cite{Gross:1981}:

\begin{equation}
\tilde{\varepsilon}_{D}(s)=\frac{\epsilon^*(s)-\epsilon_\infty}{\epsilon_0-\epsilon_\infty}=\frac{1}{1+s\sigma}
\label{debye}
\end{equation}
where $\tilde{\varepsilon}(s)_{D}$ is the complex susceptibility, $\epsilon^*$
is the complex permittivity of the dielectric, the real value $\epsilon_0$ is
the low frequency dielectric constant, the real value $\epsilon_\infty$ is
the high frequency dielectric constant and $\sigma$ is the constant
associated to the dipole's characteristic relaxation time.  

Cole and collaborators \cite{ColeCole:1941,Cole:1942} formulated a more
complete model based on experimental data on dielectrics. This new formulation
contains a parameter  $\alpha$  which can assume values between $0$ and $1$
and is given by
\begin{equation}
\tilde{\varepsilon}(s)_{CC}=\frac{1}{1+(s\sigma)^\alpha}.
\label{cce}
\end{equation}
In their studies of the dielectric relaxation process, C-D \cite{DavidsonCole:1951} 
proposed a modified equation for the dielectric
liquids glycerol and propylene-glycol. They also studied $n$-propanol and
confirmed that, in its case, the process is described by Debye's equation. The
difference from the former two liquids lies in the broadening of the dispersion
range under higher frequencies \cite{ColeCole:1941}.  

The works of C-D \cite{Davidson:1950,DavidsonCole:1951} allowed for
an almost complete determination of the dielectric properties and a more
accurate  quantitative description. The empirical expression for the complex
susceptibility $\tilde{\varepsilon}(s)$ is given by:
\begin{equation}
\tilde{\varepsilon}(s)_{CD}=\frac{1}{(1+s\sigma)^\beta},
\label{cde}
\end{equation}
where $0<\beta\leq 1$.  

In their works, H-N \cite{Negami:1966,Havril:1967} studied the
complex dielectric behavior of twenty-one polymers and noticed that
they had approximately the same form. They then arrived at an empirical
expression which generalizes the dispersion models of Debye, C-C and
C-D. This is the expression representing the relaxation process: 
\begin{equation}
\tilde{\varepsilon}(s)_{HN}=\frac{1}{(1+(s\sigma)^\alpha)^\beta}.
\label{havrinegami}
\end{equation}
It is important to observe that for $\beta=1$ we have the C-C model, while
$\alpha=1$ takes to C-D model and, finally, with $\alpha=\beta=1$ we recover
the first model proposed by Debye.
\section{Kinetic Equations}
The properties of dielectric materials are usually described by two constants
$\varepsilon'$ and $\varepsilon''$ which are called dielectric constants (or
loss factors). They can be combined in a complex dielectric constant given by  
\begin{equation}
\tilde{\varepsilon}=\varepsilon'-i\varepsilon''
\end{equation}
This constant, known as complex dielectric permittivity, is
given by the following superposition relation \cite{Frohlich:1958}:
\begin{equation}
\tilde{\varepsilon}(i\omega)=\mathscr{L} 
\left[-\frac{d\varphi(t)}{dt}\right](i\omega)=
1-i\omega\cdot\tilde{\varphi}(i\omega),
\label{permiss}
\end{equation}
where $\mathscr{L}[f(t)](i\omega)=\tilde{f}(i\omega)$ is the Laplace transform
of $f(t)$ in variable $i\omega$. We have that $\varphi(t)$ is the normalized
polarization decay function when a macroscopic electric field is removed from
its medium. Function $\varphi(t)$ contains only the contributions from the
relaxation process and we have chosen  $\varphi(0)=1$ \cite{Manning:1940}.  

In the case of linear approximation response, the polarization changes caused by thermal motion are the same as for the
macroscopic function dipole relaxation induced by the electric field \cite{Williams:1972}. Therefore, the laws governing the
dipole correlation function $\phi(t)$ are directly related to the kinetic
properties and macroscopic structures of the dielectric system, represented by
function  $\varphi(t)$. Thus, it is possible to equate the relaxation function
$\varphi(t)$ to the macroscopic dipole correlation function $\phi(t)$ as
follows:

\begin{equation}
\varphi(t)\simeq\phi(t)=\frac{\langle M(t)M(0)\rangle}{\langle M(0)M(0)\rangle},
\end{equation}
where $M(t)$ is the fluctuating macroscopic dipole moment.  

In the projection operator formalism developed by Mori \cite{Mori:1965} and
Zwanzig \cite{Zwanzig:1961}, function $\phi(t)$ is called temporal
correlation function, as the dipole correlation function defined above is a
specific case of the temporal correlation function. Thus, function $\phi(t)$,
now called temporal correlation function, has the following form in the
approximation context \cite{Boon:1980}:
\begin{equation}
\frac{d\phi(t)}{dt}=-\int_0^t K(t-x)\phi(x)dx,
\label{K}
\end{equation}
This is an integro-differential equation which takes into account the effects of
memory. Therefore, introducing the concept of an integral memory function given
by $M(t)=\int_0^t K(x)dx$ and using the fact that the relaxation function
$\varphi(t)$ also satisfies Eq.(\ref{K}), it is possible to obtain the
following relation involving function $\varphi(t)$:
\begin{equation}
\frac{d\varphi(t)}{dt}=-\frac{d}{dt}\int_0^tM(t-x)\varphi(x)dx
\equiv-\frac{d}{dt}M(t)\ast\varphi(t),
\label{convolucao}
\end{equation}
where $\ast$ denotes a convolution product.  

The relations given by Eqs.(\ref{convolucao}) and (\ref{permiss}) can now be
used to calculate the integral memory function $M(t)$. 
Applying the Laplace transform to  Eq.(\ref{convolucao}) we obtain 
\begin{equation}
\tilde{\varphi}(i\omega)=\frac{1}{i\omega(1+\tilde{M}(i\omega))},
\label{funcLaplace}
\end{equation}
where $\tilde{M}(i\omega))$ is the Laplace transform of $M(t)$.  

Substituting Eq.(\ref{funcLaplace}) into Eq.(\ref{permiss}), we obtain
the following relation:
\begin{equation}
\tilde{\varepsilon}(i\omega)=\frac{1}{1+\tilde{M}^{-1}(i\omega)}.
\label{funcaodeM}
\end{equation}

Comparing the relation expressed by Eq.(\ref{funcaodeM}) with the classical
empirical laws (\ref{debye})-(\ref{havrinegami}) (where $s=i\omega$), it is
possible to obtain the corresponding memory functions:
\begin{eqnarray}
\mbox{Debye}&\,\,\,\,\,\,&\tilde{M}_{D}(i\omega)=\displaystyle\frac{1}{i\omega\sigma},\label{MS1}\\
\mbox{C-C}&\,\,\,\,\,\,&\tilde{M}_{CC}(i\omega)=\displaystyle\frac{1}{(i\omega\sigma)^\alpha},\\
\mbox{C-D}&\,\,\,\,\,\,&\tilde{M}_{CD}(i\omega)=\displaystyle\frac{1}{(1+i\omega\sigma)^{\beta}-1},\\
\mbox{H-N}&\,\,\,\,\,\,&\tilde{M}_{HN}(i\omega)=\displaystyle\frac{1}{(1+(i\omega\sigma)^{\alpha})^{\beta}-1}.
\label{MS4}
\end{eqnarray}

Applying the inverse Laplace transform to functions (\ref{MS1})-(\ref{MS4}) we
find that the memory functions in the time variable are respectively given by:  
\begin{eqnarray}
\mbox{Debye}&\,\,\,\,\,\,&M_{D}(t)=\frac{1}{\sigma},\label{MT1}\\
\mbox{C-C}&\,\,\,\,\,\,&M_{CC}(t)=\frac{t^{\alpha-1}}{\sigma^{\alpha}\Gamma(\alpha)},\label{MT2}\\
\mbox{C-D}&\,\,\,\,\,\,&M_{CD}(t)=e^{-t/\sigma}\frac{t^{\beta-1}}{\sigma^{\beta}}E_{\beta,\beta}\left[\left(\frac{t}{\sigma}\right)^{\beta}\right],\label{MT3}\\
\mbox{H-N}&\,\,\,\,\,\,&M_{HN}(t)=\sum_{k=0}^{\infty}\left(\frac{t}{\sigma}\right)^{\alpha\beta(k+1)}t^{-1}E_{\alpha,\alpha\beta(k+1)}^{\beta(k+1)}\left[-\left(\frac{t}{\sigma}\right)^{\alpha}\right],\label{MT4}
\end{eqnarray}
with $0<\alpha\leq1$, $\,0<\beta\leq1$ and where $E_{a,b}^{c}(\cdot)$ is the
Mittag-Leffler function with three parameters. This function contains as
particular cases the Mittag-Leffler with two parameters ($c=1$) and one
parameter ($b=c=1$).  

The Mittag-Leffler functions with two and three parameters are defined,
respectively, by 
\begin{equation}
E_{a,b}(x)=\sum_{k=0}^{\infty}\frac{x^k}{\Gamma(ak+b)}
\end{equation}
and
\begin{equation}
E_{a,b}^{c}(x)=\sum_{k=0}^{\infty}\frac{(c)_k}{\Gamma(ak+b)k!}x^k.
\end{equation}
where Re($a$)$>0$, Re($b$)$>0$, Re($c$)$>0$ and $(c)_k =
\frac{\Gamma(c+k)}{\Gamma(c)}$ is the Pochhammer symbol.  Substituting 
the memory functions given by Eqs.(\ref{MT1})-(\ref{MT4}) into
Eq.(\ref{convolucao}), it is posssible to obtain the kinetic equations
associated with the models of Debye, C-C, C-D and H-N, respectively:
\begin{eqnarray}
\mbox{D}& &\frac{d\varphi(t)}{dt}+\frac{1}{\sigma}\varphi(t)=0,\label{eqcin1}\\
\mbox{C-C}& &\frac{d\varphi(t)}{dt}+\frac{1}{\sigma^{\alpha}}D_{t}^{1-\alpha}\varphi(t)=0,\\
\mbox{C-D}&&\frac{d\varphi(t)}{dt}+\frac{1}{\sigma^{\beta}}\frac{d}{dt}\left\{e^{-t/\sigma}\int_0^t(t-x)^{\beta-1}E_{\beta,\beta}^{1}\left[\left(\frac{t-x}{\sigma}\right)^{\beta}\right]e^{x/\sigma}\varphi(x)dx\right\}=0,\\
\mbox{H-N}& &\frac{d}{dt}\left\{\varphi(t)+\sum_{k=0}^{\infty}\int_0^t\frac{(t-x)^{\alpha\beta(k+1)-1}}{\sigma^{\alpha\beta(k+1)}}E_{\alpha,\alpha\beta(k+1)}^{\beta(k+1)}\left[-\left(\frac{t-x}{\sigma}\right)^{\alpha}\right]\varphi(x)dx\right\}=0.\label{eqcin4}
\end{eqnarray}
Expression $D_{t}^{\gamma}f(t)$ denotes the Riemann-Liouville fractional
derivative, defined by
\begin{equation}
D_{t}^{\gamma}f(t)=\frac{1}{\Gamma(1-\gamma)}\frac{d}{dt}\int_0^t \frac{f(x)}{(t-x)^{\gamma}}dx,\,\,\,\,\,\,\,0<\gamma\leq1.
\label{riemann}
\end{equation}
The solutions of kinetic equations (\ref{eqcin1})-(\ref{eqcin4}) are given
respectively by:
\begin{eqnarray}
\mbox{Debye}&\,\,\,\,\,\,&\varphi_{D}(t)=e^{-t/\sigma},\label{debye0}\\
\mbox{C-C}&\,\,\,\,\,\,&\varphi_{CC}(t)=E_{\alpha}\left[-\left(\frac{t}{\sigma}\right)^{\alpha}\right],\label{cc}\\
\mbox{C-D}&\,\,\,\,\,\,&\varphi_{CD}(t)=1-\left(\frac{t}{\sigma}\right)^{\beta}E_{1,\beta+1}^{\beta}\left[-\frac{t}{\sigma}\right],\label{cd}\\
\mbox{H-N}&\,\,\,\,\,\,&\varphi_{HN}(t)=1-\left(\frac{t}{\sigma}\right)^{\alpha\beta}E_{\alpha,\alpha\beta+1}^{\beta}\left[-\left(\frac{t}{\sigma}\right)^{\alpha}\right].\label{hn}
\end{eqnarray}
The graphic representation of solutions 
(\ref{debye0})-(\ref{hn}) can be seen in Figures $1$, $2$, $3$ and $4$,
respectively\footnote{In all graphs we have $\sigma=1$.}.
\begin{figure}[ht]
\begin{center}
\includegraphics[width=10cm,height=6cm]{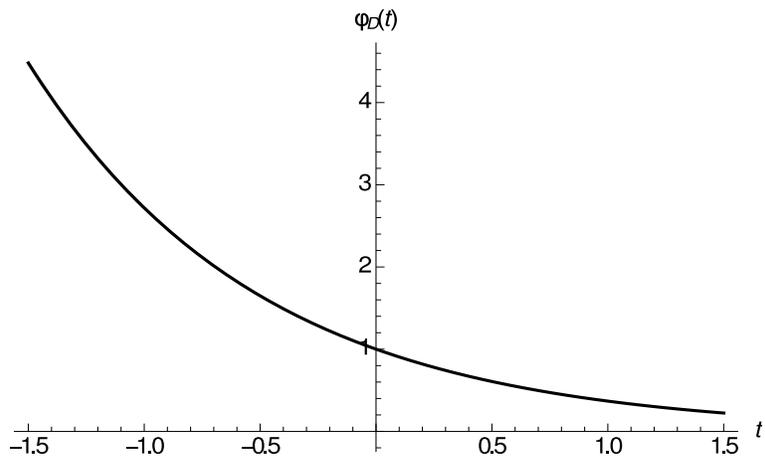}
\caption{Debye function.}
\end{center}
\end{figure}
\begin{figure}[ht]
\begin{center}
\includegraphics{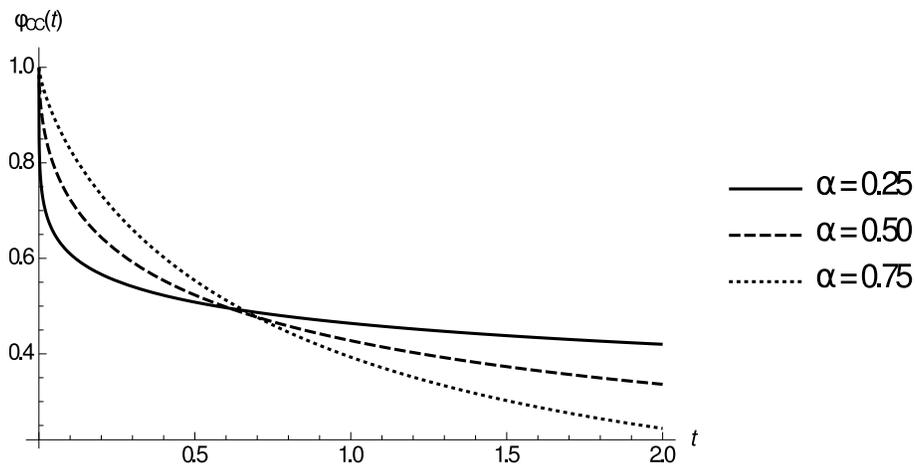}
\caption{Cole-Cole function.}
\end{center}
\end{figure}
\newpage \clearpage
\begin{figure}[ht]
\begin{center}
\includegraphics{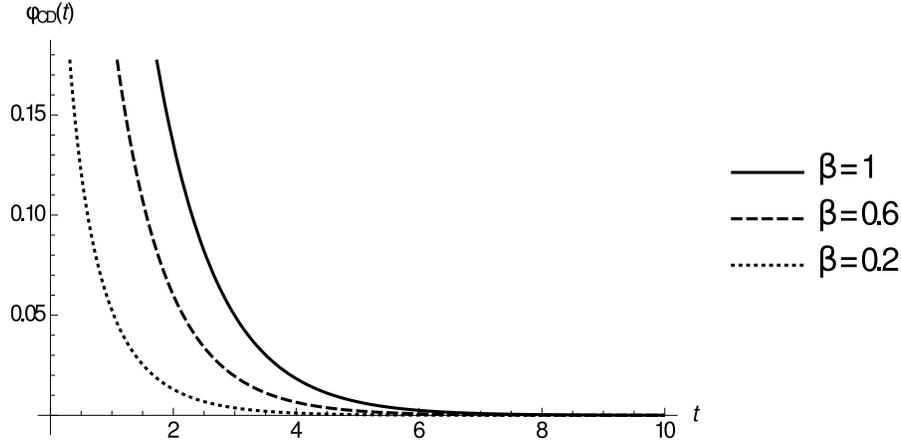}
\caption{Cole-Davidson function.}
\end{center}
\end{figure}
\begin{figure}[ht]
\begin{center}
\includegraphics{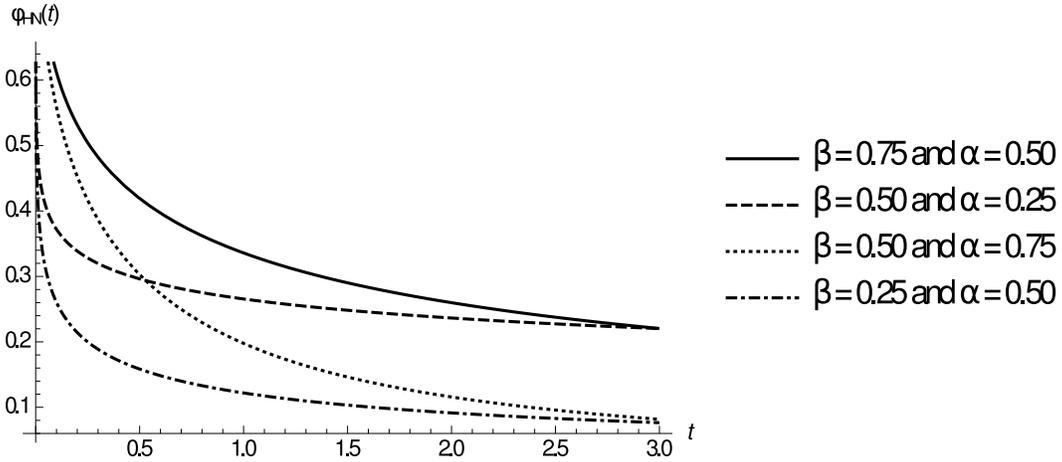}
\caption{Havriliak-Negami function.}
\end{center}
\end{figure}

\section{Fractional Kinetic Equations}

As mentioned above, the generalization of the laws describing anomalous
relaxation phenomena in dielectrics has been the object of studies by several
important researchers. Here we turn our attention to a recent study
\cite{Capelas:2014} in which the following initial value problem has been
considered:

\begin{equation}
 \frac{d^\alpha}{dt^\alpha}u(t)=
-\lambda t^{\beta}u(t),\,\,\,\,\,\,t>0,\,\,\,\,\,u(0)=u_0, 
\label{kohl}
\end{equation}
with $u_0$ and $\lambda$ positive constants. Parameters $\alpha$ and
$\beta$ are subject to the conditions

\begin{equation}
0<\alpha\leq1\,\,\,\,\mbox{and}\,\,\,\,-\alpha<\beta\leq1-\alpha .
\label{condi}
\end{equation}
Fractional equation (\ref{kohl}) appears as a generalization of the
well-known model by Kohlrausch (Kohlrausch-Willians-Watts) based on the
``stretched'' exponential function
\cite{Williams:1970,Anderssen:2004,Kohlrausch:1854} and its solution is
represented in terms of a function introduced by Kilbas and Saigo
\cite{Kilbas:1995,Saigo:1995}, which is given by the series

\begin{equation}
E_{\alpha,m,l}=\sum_{n=0}^{\infty}c_n z^n,\,\,\,\,c_n=
\prod_{i=0}^{n-1}\frac{\Gamma[\alpha(im+l)+1]}{\Gamma[\alpha(im+l+1)+1]}
\label{sol}
\end{equation}
with $\alpha,m, l\in \mathbb{R}$,
$\alpha>0$, $m>0$ and $\alpha(im+l)\neq-1,-2,-3,\ldots$  
Without loss of generality, we can take $\lambda=1$  in problem (\ref{kohl});
the solution is then 

\begin{equation}
u(t)=E_{\alpha,1+\frac{\beta}{\alpha}\frac{\beta}{\alpha}}(-t^{\alpha+\beta})
=1+\sum_{n=1}^{\infty}(-1)^n\prod_{i=0}^{n-1}
\frac{\Gamma(i(\alpha+\beta)+\beta+1)}{\Gamma(i(\alpha+\beta)+\alpha+\beta+1)}
t^{(\alpha+\beta)n}
\end{equation}
The conditions given in Eq.(\ref{condi}) guarantee the existence and complete
monotonicity of solution $u(t)$ for $t\geq0$. We now mention some particular cases.  

For $\beta=0$ we have the particular case $u(t)= E_{\alpha,1,0} (-t^{\alpha})=
E_{\alpha}(-t^{\alpha})$, where $E_{\alpha}(\cdot)$ is the classical 
Mittag-Leffler function \cite{Gorenflo:2014}. The second particular case is associated 
with the well-known ``stretched'' exponential function and is given by $u(t)=E_{1,1+\beta,\beta}(-t^{\beta+1})
=\exp\left( -\frac{t^{\beta+1}}{\beta+1}\right)$ where $\alpha=1$ and
$-1<\beta\leq0$.  

Still supposing $\beta=0$ one finds the solution $u(t)=E_{1,1,0}(-t)=\exp(-t)$,
which is the exponential solution to the relaxation equation.  With an analogous generalization purpose, 
Garra et al.\ \cite{Mainardi:2014}, using the general theory of the hyper-Bessel operator \cite{Kiryakova:1994},
solved a generalized relaxation equation which was called fractional differential equation with time-variant 
coefficient. In that equation, operator $\left(t^{\theta}\frac{d}{dt}\right)^{\alpha}$ replaces the
derivatives in the relaxation equation.  

Here, the aim of the paper, we will consider the following fractional differential equation:

\begin{equation}
D_{t}^{\gamma}\varphi(t)=-D_{t}^{\gamma}
\left\{M(t)\ast\varphi(t)\right\}, \qquad \mbox{with} \quad \gamma \in (0,1] . 
\label{fracionaria}
\end{equation}
In this equation, operator $D_{t}^{\gamma}$ is the Riemann-Liouville fractional
derivative, defined in Eq.(\ref{riemann}). Imposing the normalization condition
$D^{\gamma-1}_{t}\varphi(0)=1$, it is possible to obtain the fractional kinetic
equations with the help of the memory functions explained in the 
previous section.

First, memory function Eq.(\ref{MT1}) is substituted into Eq.(\ref{fracionaria})
in order to obtain:
\begin{equation}
D^{\gamma}_{t}\varphi(t)=- D^{\gamma}_{t}\left\{M_{D}\ast\varphi(t)\right\}=- D^{\gamma}_{t}\left\{\frac{1}{\sigma}\ast\varphi(t)\right\}.
\label{debyef1}
\end{equation}
Applying the Laplace transform to it, we obtain 
\begin{equation}
s^{\gamma}\tilde{\varphi}(s)-1=-s^{\gamma}\frac{\tilde{\varphi}(s)}{s\sigma},\,\,\,\,\,\,\,\,\,\,\mbox{Re($s$)}>0.
\end{equation}
Then, isolating $\tilde{\varphi}(s)$ we get
\begin{equation}
\tilde{\varphi}(s)=\frac{s^{1-\gamma}}{s+\sigma^{-1}}.
\label{lf1}
\end{equation}
Calculating its inverse Laplace transform, Debye's fractional relaxation
function turns out to be

\begin{equation}
\varphi_{DF}(t)= t^{\gamma-1}E_{1,\gamma}\left(-\frac{t}{\sigma}\right).
\label{debyef2}
\end{equation}
For $\gamma=1$ we recover Debye's exponential solution 
$\varphi_{D}(t)= e^{-t/\sigma}$.  \\
In the same way, substituting Eq.(\ref{MT2}) into Eq.(\ref{fracionaria})
we have
\begin{equation}
D^{\gamma}_{t}\varphi(t)=- D^{\gamma}_{t}\left\{M_{CC}\ast\varphi(t)\right\}=- D^{\gamma}_{t}\left\{\frac{t^{\alpha-1}}{\sigma^{\alpha}\Gamma(\alpha)}\ast\varphi(t)\right\}.
\label{ccf1}
\end{equation}
Again, applying Laplace transform to Eq.(\ref{ccf1}) we obtain the following 
expression:
\begin{equation}
\varphi(s)=\frac{s^{\alpha-\gamma}}{s^\alpha+\sigma^{-\alpha}},
\label{lf2}
\end{equation}
whence emerges the C-C fractional relaxation function:
\begin{equation}
\varphi(t)_{CCF}= t^{\gamma-1}E_{\alpha,\gamma}
\left[-\left(\frac{t}{\sigma}\right)^\alpha\right].
\label{ccf2}
\end{equation}
For $\alpha=1$ we recover Debye's fractional function Eq.(\ref{debyef2}).  For
$\gamma=1$ one has the C-C model given by Eq.(\ref{cc}).  

In order to generalize the C-D model, Eq.(\ref{fracionaria}) can 
be substituted into  Eq.(\ref{MT3}), in order to obtain

\begin{equation}
D^{\gamma}_{t}\varphi(t)=- D^{\gamma}_{t}\left\{M_{CD}\ast\varphi(t)\right\}=
- D^{\gamma}_{t}\left\{\left[e^{-\frac{t}{\sigma}}\sigma^{-\beta}t^{\beta-1}E_{\beta,\beta}\left[-\left(\frac{t}{\sigma}\right)^{\beta}\right]\right]\ast\varphi(t)\right\}.
\label{cdf1}
\end{equation}
Applying the Laplace transform to the last expression, we obtain:

\begin{equation}
s^{\gamma}\tilde{\varphi}(s)-1=-s^{\gamma}\tilde{\varphi}(s)\mathscr{L}\left\{e^{-\frac{t}{\sigma}}
\sigma^{-\beta}t^{\beta-1}E_{\beta,\beta}\left[-\left(\frac{t}{\sigma}\right)^{\beta}\right]\right\}(s).
\end{equation}
Expanding the exponential function and the Mittag-Leffler function with two
parameters in power series, we get

\begin{equation}
s^{\gamma}\tilde{\varphi}(s)-1=-s^{\gamma}\tilde{\varphi}(s)\mathscr{L}\left[\sum_j^{\infty}
\left(-\frac{t}{\sigma}\right)^j\frac{1}{j!}\frac{t^{\beta-1}}{\sigma^{\beta}}\sum_k^{\infty}\left(\frac{t}{\sigma}\right)^{\beta k}\frac{1}{\Gamma(\beta k +\beta)}\right](s).
\end{equation}
Multiplying and dividing by $\Gamma(\beta+ j+\beta k)$ and rearranging factors,
this gives:
\begin{equation}
s^{\gamma}\tilde{\varphi}(s)-1=-s^{\gamma}\tilde{\varphi}(s)\mathscr{L}\left[\sum_k^{\infty}t^{\beta k+\beta -1}E_{1,\beta k+\beta}^{\beta k+\beta}\left(-\frac{t}{\sigma}\right)\frac{1}{\sigma^{\beta k+\beta}}\right](s)
\end{equation}
which results in
\begin{equation}
s^{\gamma}\tilde{\varphi}(s)-1=
-\frac{s^{\gamma}\tilde{\varphi}(s)}{(1+\sigma s)^\beta-1},
\end{equation}
or, isolating $\tilde{\varphi}(s)$,
\begin{equation}
\tilde{\varphi}(s)=\frac{1}{s^{\gamma}}-\frac{s^{-\gamma}}{(1+s\sigma)^{\beta}}.
\label{lf3}
\end{equation}
Thus, using the inverse Laplace transform we finally get 
\begin{equation}
\varphi(t)_{CDF}= \frac{t^{\gamma-1}}{\Gamma(\gamma)}-\sigma^{-\beta}t^{\beta+\gamma-1}E_{1,\beta+\gamma}^{\beta}\left(-\frac{t}{\sigma}\right)
\label{cdf2}
\end{equation}
Now, taking $\beta=1$ in Eq.(\ref{cdf2}) and using the following relation
involving the Mittag-Leffler function with two parameters: 

\begin{equation}
E_{\alpha,\gamma}(y)=\frac{1}{\Gamma(\gamma)}+yE_{\alpha,\alpha+\gamma}(y) , 
\label{ML2}
\end{equation}
we recover the function of Debye's fractional model, given by
Eq.(\ref{debyef2}). For $\gamma=1$ we recover the C-D function,
given by Eq.(\ref{cd}).

Finally, we substitute H-N's memory function, given by Eq.(\ref{MT4}), in Eq.(\ref{fracionaria}), we have: 
\begin{equation}
D^{\gamma}_{t}\varphi(t)=- D^{\gamma}_{t}\left\{M_{HN}\ast\varphi(t)\right\}=- D^{\gamma}_{t}\left\{\left[\sum_{k=0}^{\infty}
\sigma^{-\alpha k(\beta+1)}t^{\alpha\beta(k+1)-1}E_{\alpha,\alpha\beta(k+1)}^{\beta(k+1)}\left[-\left(\frac{t}{\sigma}\right)^{\alpha}\right]\right]\ast\varphi(t)\right\}.
\label{hnf1}
\end{equation}
Applying the Laplace transform and using the following equality given in \cite{Capelas:2011},

\begin{equation}
\mathscr{L}[t^{\gamma-1} E_{\alpha,\gamma}^{\beta}(at^{\alpha})](s)= \frac{s^{\alpha \beta-\gamma}}{(s^{\alpha}-a)^\beta},
\end{equation}
we obtain
\begin{equation}
s^{\gamma}\tilde{\varphi}(s)-1
=-s^{\gamma}\tilde{\varphi}(s)
\frac{1}{(1+(s\sigma)^{\alpha})^{\beta}}\sum_{k=0}^{\infty}
\frac{1}{(1+(s\sigma)^{\alpha})^{\beta k}}
\end{equation}
or, equivalently, 
\begin{equation}
s^{\gamma}\tilde{\varphi}(s)-1=-\frac{s^{\gamma}\tilde{\varphi}(s)}{(1+(s\sigma)^{\alpha})^{\beta}}\,\cdot\,\frac{1}{1-(1+(s\sigma)^{\alpha})^{-\beta}},
\end{equation}
From this expression it follows that
\begin{equation}
\tilde{\varphi}(s)=\frac{1}{s^{\gamma}}-\frac{s^{-\gamma}}{(1+(s\sigma)^{\alpha})^{\beta}}.
\label{lf4}
\end{equation}
Applying the inverse Laplace transform, we obtain
\begin{equation}
\varphi(t)_{HNF}= \frac{t^{\gamma-1}}{\Gamma(\gamma)}-\sigma^{-\alpha\beta}t^{\alpha\beta+\gamma-1}E_{\alpha,\alpha\beta+\gamma}^{\beta}\left[-\left(\frac{t}{\sigma}\right)^\alpha\right].
\label{hnf2}
\end{equation}
If $\alpha=1$ in Eq.(\ref{hnf2}), we recover the function of the fractional
C-D model Eq.(\ref{cdf2}). If we put $\beta=1$ in Eq.(\ref{hnf2}), it is
then possible to recover the function of the fractional C-C model given
by Eq.(\ref{ccf2}). Finally, for $\alpha=\beta=1$ we have function of the
fractional Debye model, Eq.(\ref{debyef2}). The H-N model Eq.(\ref{hn}) is recovered when $\gamma=1$.  

Therefore, we have just shown that the models by Debye, C-C, C-D and H-N given by Eqs.(\ref{debye0}), (\ref{cc}),
(\ref{cd}) and (\ref{hn}) can be considered particular cases of the
fractional models given by Eqs.(\ref{debyef2}), (\ref{ccf2}), (\ref{cdf2}) and
(\ref{hnf2}), respectively. 

The complex dielectric permittivity (in variable $s$) is given in the
fractional case by the following superposition relation:
\begin{equation}
\tilde{\varepsilon}(s)=\mathscr{L}\left[- D^{\gamma}_{t}\varphi(t)\right](s)=1-s^{\gamma}\tilde{\varphi}(s).
\label{complexafrac}
\end{equation}
With this definition, we recover in the four fractional relaxation function
models the corresponding classical empirical responses described by
Eqs.(\ref{debye}), (\ref{cce}), (\ref{cde}) and (\ref{havrinegami}). To see
this we need only to substitute Eqs.(\ref{lf1}), (\ref{lf2}), (\ref{lf3})
and (\ref{lf4}) into Eq.(\ref{complexafrac}).  

Figures $5$, $6$, $7$ and $8$ graphically represent the solutions given by
Eqs.(\ref{debyef2}), (\ref{ccf2}), (\ref{cdf2}) and (\ref{hnf2}), respectively.

\begin{figure}[ht]
\begin{center}
\includegraphics{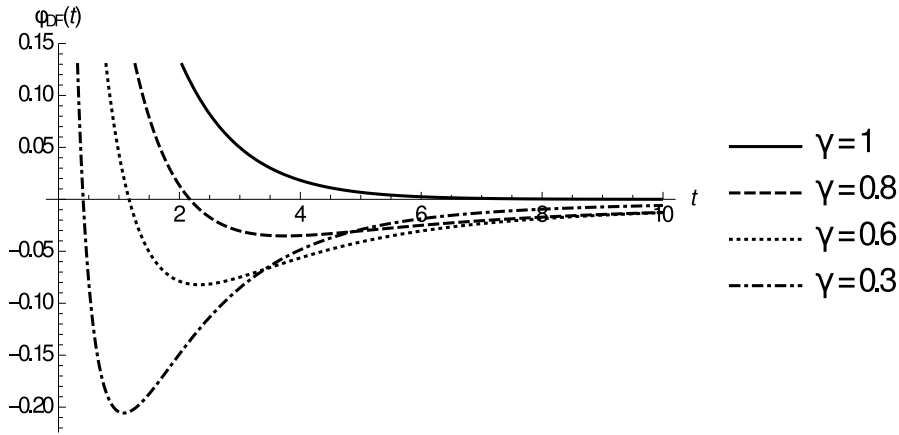}
\caption{Fractional Debye function.}
\end{center}
\end{figure}
\begin{figure}[ht]
\begin{center}
\includegraphics{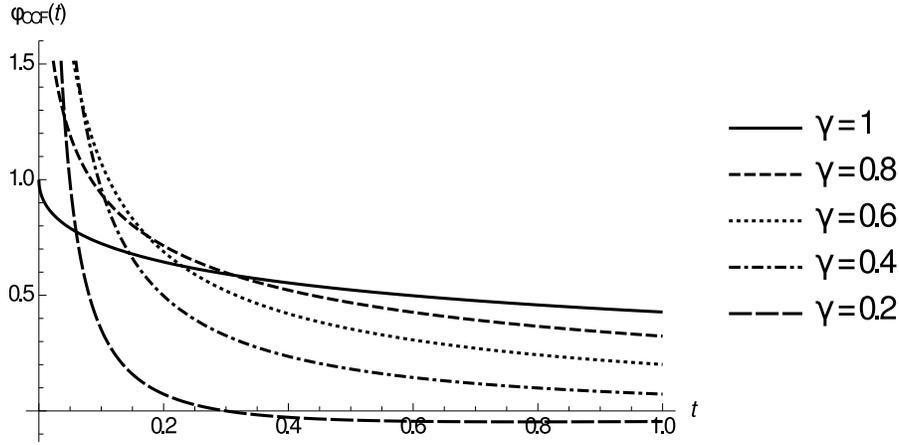}
\caption{Fractional Cole-Cole function with $\alpha=0.5$.}
\end{center}
\end{figure}
\begin{figure}[ht]
\begin{center}
\includegraphics{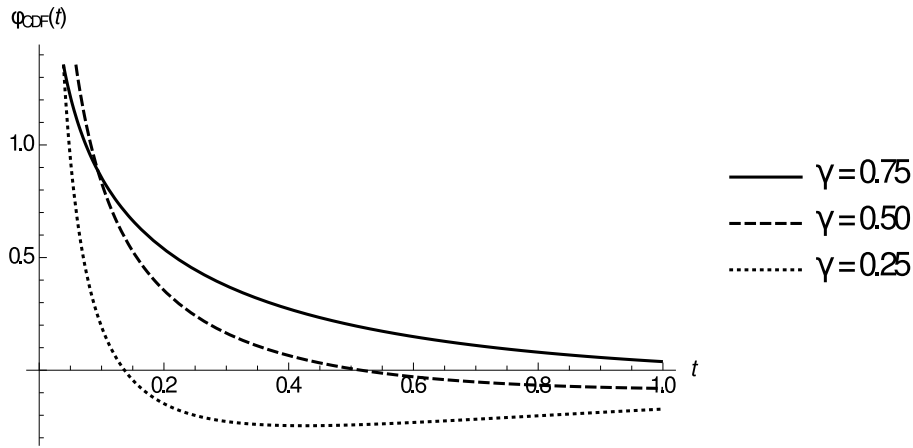}
\caption{Fractional Cole-Davidson function with $\beta=0.5$.}
\end{center}
\end{figure}
\begin{figure}[ht]
\begin{center}
\includegraphics{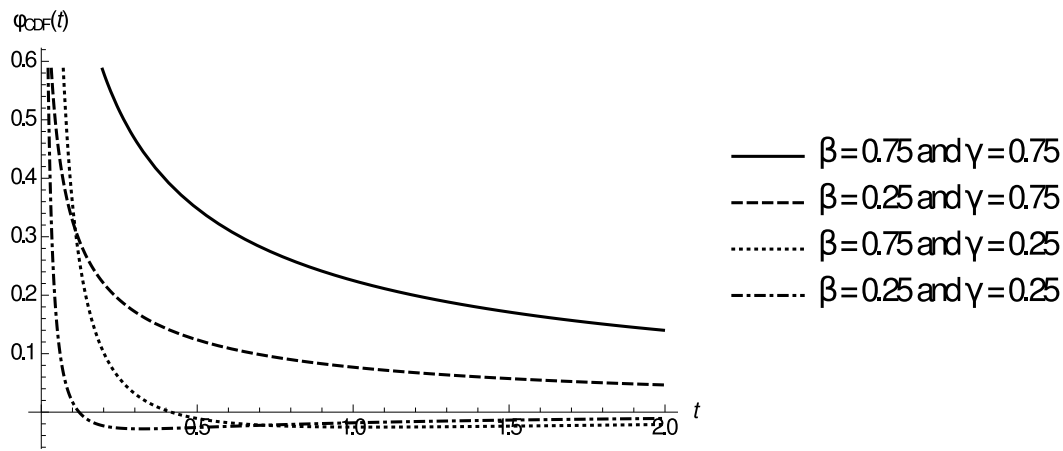}
\caption{Fractional Havriliak-Negami function with $\alpha=0.5$.}
\end{center}
\end{figure}
 
\section{Conclusion}
 
This study has shown that certain fractional kinetic equations with a parameter
$0<\gamma\leq1$, whose solutions are given in terms of Mittag-Leffler
functions, can be used as generalized models of the classical relaxation
kinetic equations. The classical solutions turned out to correspond to the
particular case $\gamma=1$.   

In all the cases discussed, the results of the corresponding classical models
(response functions) were recovered, as well as the empirical dielectric
complex functions (in variable $s$).

The graphs of the  fractional relaxation functions and of the solutions of the
fractional kinetic equations were constructed. The analysis of fractional
relaxation function graphs and their algebraic expressions given by
Eqs.(\ref{debyef2}), (\ref{ccf2}), (\ref{cdf2}) and (\ref{hnf2}) has shown that
the functions's signals and their complete monotonicity depend on the
parameters involved. These aspects deserve a more careful analysis and will be
the subject of a future work \cite{Ester:2015}.

\section*{Acknowledgment}
We are indebted to Dr. J. Em\'{\i}lio Maiorino for several and usefull discussions.
(ECFAR) thanks CAPES for a research grant.


\end{document}